\documentclass[preprint,preprintnumbers,amsmath,amssymb]{revtex4}

\usepackage{graphicx}

\usepackage{comment}
\usepackage[normalem]{ulem}

\renewcommand{\thefigure}{\arabic{figure}}

\makeatletter
\def\blfootnote{\xdef\@thefnmark{}\@footnotetext}
\makeatother

\usepackage{color}
\usepackage{ulem}

\begin{document}
\renewcommand{\figurename}{\textbf{Fig.}}
\renewcommand{\thefigure}{\textbf{\arabic{figure}}}
%
\title{Quantum Oscillations of Electrical Resistivity in an Insulator}

\author{Z. Xiang$^{1}$, Y. Kasahara$^2$, T. Asaba$^1$, B. Lawson$^{1,3}$, C. Tinsman$^1$, Lu Chen$^1$, K. Sugimoto$^4$, S.\,Kawaguchi$^4$, Y. Sato$^2$, G. Li$^{1}$, S. Yao$^5$, Y. L. Chen$^6$, F. Iga$^7$, John Singleton$^8$}
 \author{Y. Matsuda$^2$}
 \email{matsuda@scphys.kyoto-u.ac.jp}
\author{Lu Li$^{1}$}
 \email{luli@umich.edu}

\affiliation{
$^1$Department of Physics, University of Michigan, Ann Arbor, MI  48109, USA\\
$^2$Department of Physics, Kyoto University, Kyoto 606-8502, Japan\\
$^3$Faculty of Applied Science, Universit\'e Chr\'etienne Bilingue du Congo, Beni, North-Kivu, Democratic Republic of Congo\\
$^4$Japan Synchrotron Radiation Research Institute, Sayo, Hyogo 679-5198, Japan\\
$^5$National Laboratory of Solid State Microstructures, Nanjing University, Nanjing 210093, China\\
$^6$Department of Physics, Clarendon Laboratory, University of Oxford, Oxford, OX1 3PU, UK\\
$^7$College of Science, Ibaraki University, Mito 310-8512, Japan\\
$^8$National High Magnetic Field Laboratory, Los Alamos National Laboratory, Los Alamos,  NM 87545
}

\date{\today}
\begin{abstract}
In metals, orbital motions of conduction electrons on the Fermi surface are quantized in magnetic fields, which is manifested by quantum oscillations in electrical resistivity. This Landau quantization is generally absent in insulators. Here we report a notable exception in an insulator --- ytterbium dodecaboride (YbB$_{12}$). Despite much larger than that of metals,the resistivity of YbB$_{12}$ exhibits profound quantum oscillations. This unconventional oscillation is shown to arise from the insulating bulk, yet the temperature dependence of their amplitude follows the conventional Fermi liquid theory of metals. The large effective masses indicate the presence of Fermi surface consisting of strongly correlated electrons. Our result reveals a mysterious bipartite ground state of YbB$_{12}$: it is both a charge insulator and a strongly correlated metal.

\end{abstract}


\maketitle                   

Strong electronic correlations bring novel physical phenomena. A remarkable example is the unique ground state of Kondo insulators, in which the hybridization between itinerant and localized electrons leads to the opening of an insulating  gap and consequently a divergent low-temperature resistance \cite{Kondolattice,riseborough}. Recently the nature of the ground state of a Kondo insulator in intense magnetic fields has become a big mystery in condensed matter physics.  In mixed-valence compound samarium hexaboride (SmB$_6$), although the sample is a good insulator and its resistance increases by five orders of magnitude during cooling-downs, Landau level (LL) quantization happens under applied magnetic field, and strong quantum oscillations are observed in magnetization (the de Haas-van Alphen, or dHvA,  effect) \cite{liSmB6,tan2015unconventional}.  Origin and interpretation of the dHvA oscillations in SmB$_6$ have been  highly controversial, as there are a number of peculiar features. Firstly, the oscillations are observed only in magnetization, and not in electrical resistivity (the Shubnikov-de Haas, or SdH, effect).  Secondly, despite the strong electron correlations, only the oscillations arising from quasiparticles having very light effective masses ($m\ll m_e$, $m_e$ is the free electron mass) are observed. Thirdly, in floating-zone grown SmB$_6$ samples the dHvA signal exhibits a striking deviation from the standard Lifshitz-Kosevich (LK) formula of the Fermi-liquid theory \cite{tan2015unconventional}. These observations point to either the topologically protected surface state \cite{liSmB6}, or an unconventional Fermi surface in insulators \cite{tan2015unconventional}.  A number of intriguing physical origins have been proposed, including exciton-based magnetic breakdown \cite{knolle2015,knolle2017}, Majorana type charge-neutral Fermi surfaces \cite{baskaran2015majorana}, and a failed superconducting ground state \cite{erten2017skyrme}.  Recent theories suggest that an emergent neutral Fermi sea can exist in a mixed-valence gapped system, which may exhibit both dHvA and SdH oscillations under external magnetic field \cite{Senthil2017neutralFS,Senthil2017QO}. A key to solving the most fundamental problem, ``the Fermi surface in an insulator", lies in the clarification whether the quantum oscillations, in particular in the charge transport, are observable in another insulating system.

\begin{figure}[t]
\centering
\includegraphics[width=0.6\columnwidth]{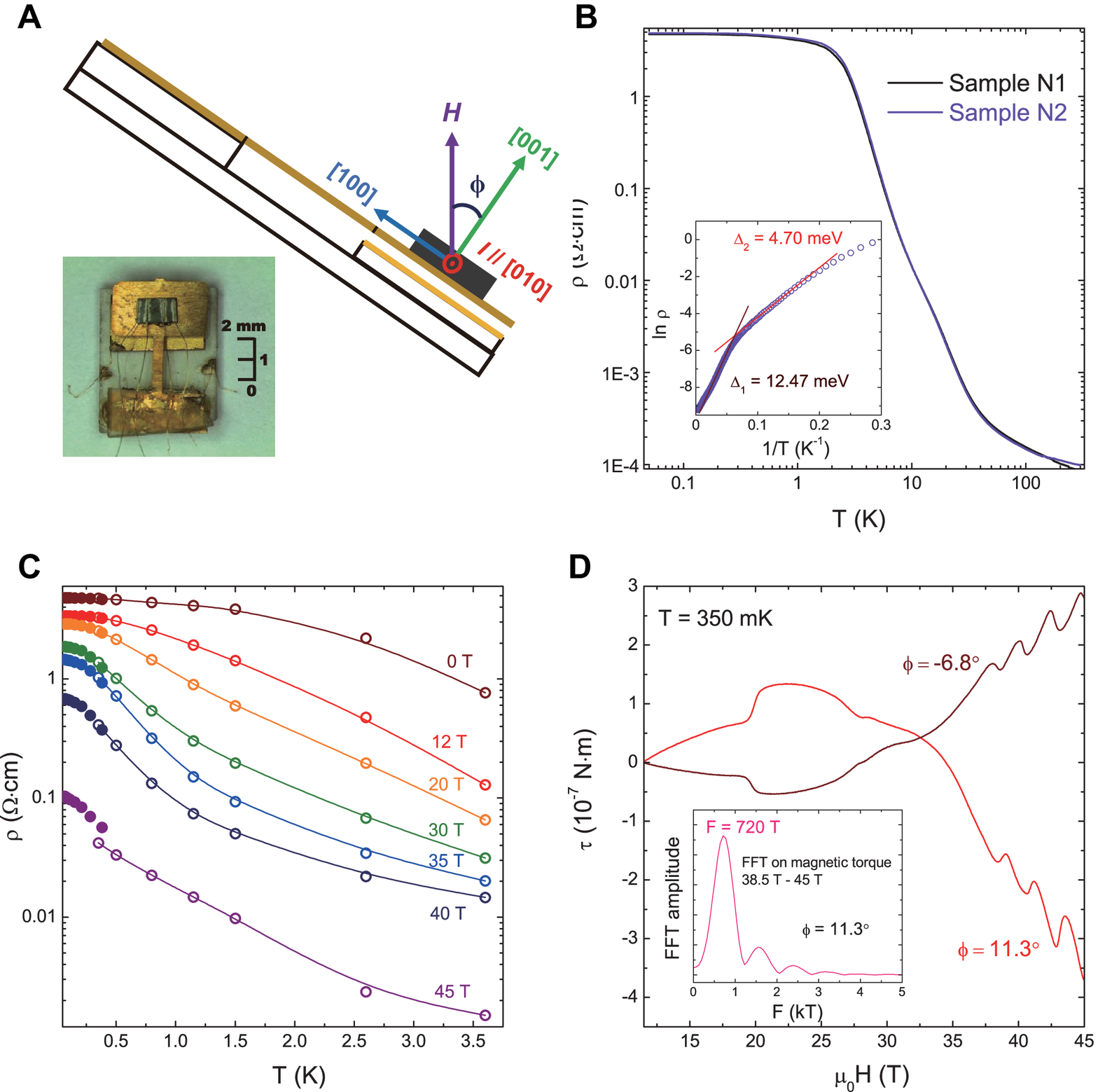}
\caption{{\bf Experimental setup of electrical transport and magnetic torque measurements in YbB$_{12}$.}
(\textbf{A}) Sketch of the experimental setup (details see \cite{SM}) and the definition of tilt angle $\phi$ with respect to the magnetic field. Inset: Photograph of a YbB$_{12}$ single crystal (Sample N2) mounted onto a cantilever beam magnetometer, with four gold wires attached to the crystalline (001) surface for the transport measurement.
(\textbf{B}) Resistivity of two YbB$_{12}$ single crystals plotted as a function of temperature. The inset shows the Arrhenius plot, $\ln \rho$ vs. 1/$T$. According to the thermal activation model, the slope of the Arrhenius plot equals to $\Delta/2k_B$, with $\Delta$ the bandgap width and $k_B$ the Boltzmann constant. Linear fitting in two different temperature range, 20 K $< T <$ 40 K and 6 K $< T <$ 12.5 K, yields two characteristic gap width 12.5 meV and 4.7 meV, respectively.
(\textbf{C}) Resistivity of YbB$_{12}$ sample N2 under different magnetic fields from 0 T to 45 T, plotted against temperature. Hollow and solid symbols are data taken in $^3$He cryostat at $\phi$ = 7.4$^\circ$ and in dilution fridge at $\phi$ = 8.5$^\circ$, respectively. Solid lines are guides to the eye.
(\textbf{D}) Magnetic torque in YbB$_{12}$ measured at $T$ = 350 mK and at two different tile angles, $\phi$ = -6.8$^\circ$ and $\phi$ = 11.3$^\circ$. Both exhibit strong quantum oscillations under high magnetic field. The oscillatory torque at $\phi$ = 11.3$^\circ$ reaches an amplitude $\sim$ 6$\times$10$^{-8}$ N$\cdot$m at the highest field, corresponding to an effective transverse magnetization of $\sim$ 1.4$\times$10$^{-9}$ A$\cdot$m$^2$ (1.51$\times$10$^{14}$ $\mu_B$).  Inset: FFT on the magnetic torque signal with $\phi$ = 11.3$^\circ$ reveals a major peak at $F$ = 720 T and its harmonics.
}
\label{FigData}
\end{figure}

\begin{figure}[t]
\centering
\includegraphics[width=1.0\columnwidth]{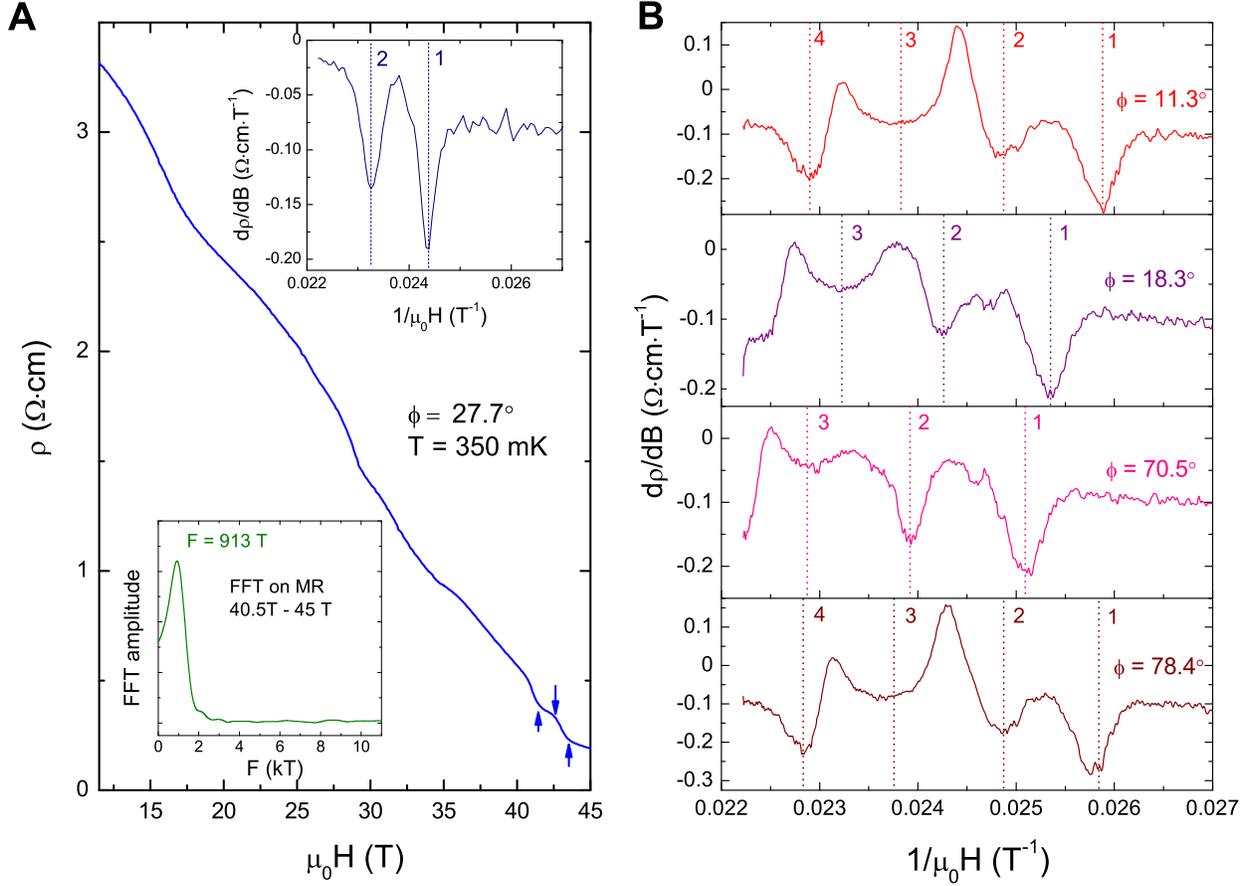}
\caption{{\bf Resistivity and electrical oscillations in intense magnetic fields in YbB$_{12}$.}
(\textbf{A}) Resistivity as a function of magnetic field measured up to 45 T taken at $T$ = 350 mK at a tilt angle $\phi$ = 27.7$^\circ$. Quantum oscillations are clearly observed at high magnetic field beyond 40.8 T. The extrema are marked by arrows. Upper inset, First magnetic field derivative of resistance is shown in the inset with two notable valleys and one peak. Lower inset, FFT on the magnetoresistance data presented between $\mu_0H$ = 40.5 T and 45 T. A single peak frequency of $F$ = 913 T is resolved.
(\textbf{B}) Field derivative of sample resistivity at different tilt angles. Dotted lines mark the approximately evenly spaced valleys of SdH oscillation. Three to four periods in total can be observed, depending on the field orientation. The oscillation pattern at $\phi$ is generally repeated at 90$^\circ-\phi$, consistent with the cubic symmetry of crystal structure.
}
\label{FigQO}
\end{figure}

\begin{figure*}[htbp!]
\centering
\includegraphics[width=0.95\textwidth]{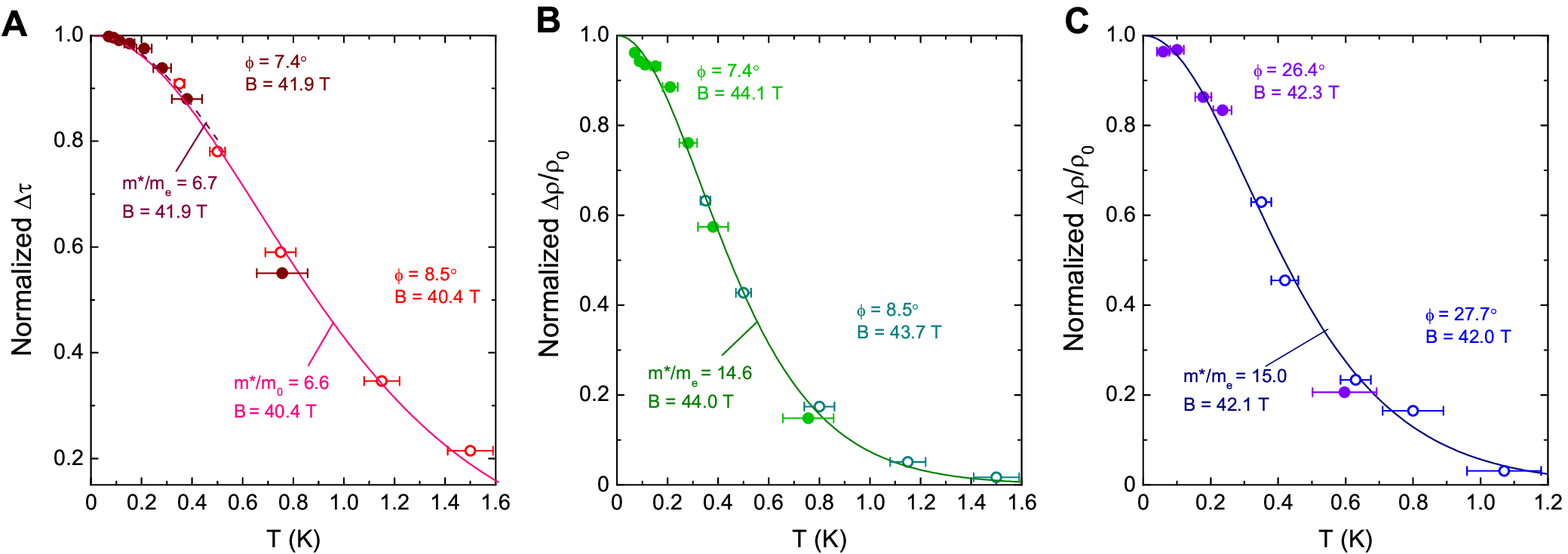}
\caption{{\bf Effective mass fitting on dHvA and SdH oscillations in YbB$_{12}$.} (\textbf{A}) Normalized dHvA oscillation amplitude $\Delta \tau$ as a function of temperature. Hollow circles and solid circles are the data taken in $^3$He cryostat at $\phi$ = 8.5$^\circ$ and in portable dilution fridge at $\phi$ = 7.4$^\circ$, respectively. LK model simulation using parameters $m^*$ = 6.6 (6.7) $m_e$, $B$ = 40.4 (41.9) T is presented by solid (dashed) line. Temperature dependence of normalized SdH amplitude $\Delta \rho/\rho_0$ at (\textbf{B}) the same tilt angles as in (\textbf{A}), and (\textbf{C}) tilt angle close to that in Fig. 2A. Hollow and solid symbols are data measured in portable $^3$He cryostat and dilution fridge, respectively. Here $\rho_0$ is the zero-field resistivity at corresponding temperatures. Solid lines are simulations based on LK formula, with parameters shown in each panels. According to the simulations, SdH effective mass exhibits a weak SdH mass anisotropy between $\phi \simeq$ 8$^\circ$ ((\textbf{B}), $m^*$ = 14.6 $m_e$) and $\phi \simeq$ 27$^\circ$ ((\textbf{C}), $m^*$ = 15.0 $m_e$). In all three panels, the quantum oscillation amplitudes (raw data shown in Fig. S3 in \cite{SM}) are taken as the averaged value of a pair of adjacent peak and valley obtained after subtracting a polynomial background from the raw data, and the effective magnetic field labeled in each panel is an inverse average of peak position and valley position. The error bars on temperature are estimated based on the magnetoresistance effect on the reading of the ruthenium oxide thermometer above 11.4 T.
}
\label{FigMass}
\end{figure*}

We chose to study a different mixed-valence Kondo insulator, YbB$_{12}$, another cubic-structured rare-earth intermetallic compound possessing strong interactions between localized \textit{4f} and itinerant \textit{5d} orbitals  \cite{Kasaya1983,Kasaya1985,Kasuya}. Early theories predicted an existence of topological surface states due to mirror-symmetry protection in YbB$_{12}$ \cite{TCPYbB12}, which makes it a topological crystalline Kondo insulator. YbB$_{12}$ single crystals are grown by floating zone method (section S1 in \cite{SM}).  Using the experimental setup illustrated in Fig.\,1A, magnetic torque and magnetoresistance (MR) are measured up to 45\,T simultaneously (section S2 in \cite{SM}). The temperature dependence of the resistivity shown in Fig.\,1B confirms an increase of five orders of magnitude from room temperature to 50\,mK. A weakly-temperature-dependent resistivity appears below 2.2 K, resembling the resistive ``plateau" well known in SmB$_6$ at $T <$ 3.5 K \cite{SmB6pressure}, which is an indication of the existence of extended in-gap states. Fitting with the thermal activation model of resistivity, $\rho(T)$ = $\rho_0$$\exp(\Delta/2k_BT)$, reveals a two-gap feature with the gap width 12.5 meV (20 K $< T <$ 40 K) and 4.7 meV (6 K $< T <$ 12.5 K), respectively (inset of Fig.\,1B), consistent with former transport result \cite{IgaGrowth}. Upon applying the magnetic field, the negative slope of $\rho(T)$ curve is preserved up to 45\,T with no hints of metallic behaviour (Fig.\,1C),  indicating that the ground state is still gapped.


Field dependence of the magnetic torque in the insulating state of YbB$_{12}$ is shown in Fig.\,1D.  Subtle metamagnetic transitions and/or crossover can be seen at 20\,T and 27\,T,  potentially related to the predicted field-induced staggered magnetism in Kondo insulator \cite{PatrickLee}.  Above approximately 37\,T, the dHvA oscillations is clearly resolved (Fig.\,1D).  We note that the dHvA oscillations appear well below the magnetic-field-driven gap closing, i.e., the insulator-metal transition in YbB$_{12}$ \cite{IgaPulsedField} (for details see section S3 and Fig . S1 in \cite{SM}).
Fast Fourier Transform (FFT) on the $\phi$ = 11.3$^\circ$ torque curve gives a dHvA frequency of $F$ = 720 T (inset of Fig.\,1D).

In Fig.\,2A the MR data at $\phi$ = 27.7$^\circ$  is plotted between 11.5\,T and 45\,T. Given the zero-field resistivity $\rho(0)$ = 4.67 $\Omega\cdot$cm of this sample,  a significant negative MR ($\{\rho(H)-\rho(0)\}/\rho(0)$)  of -95.9$\%$ is achieved at 45 T (A detailed angular dependence of MR is shown in Fig. S2 in \cite{SM}).  The negative MR is a hallmark in Kondo insulators as a result of the field suppression of the hybridization gap \cite{Sugiyama,CeNiSn,Boebinger,Cooley}. We note that the negative MR in YbB$_{12}$ is much larger then that in SmB$_6$ \cite{SmB6pulsed,wolgast2015magnetotransport}. This is probably due to the larger effective Land\'{e} $g$-factor and the more localized nature of $f$ electrons in YbB$_{12}$, which lead to the stronger impact of magnetic field \cite{Cooley,IgaPulsedField}.

The most striking result in Fig.\,2A is the oscillating wiggles appear in the MR under strong magnetic fields. From 40.8 T up to 45 T, two valleys and one peak in total can be clearly observed (upper inset of Fig.\,2A), and the FFT on MR in such a field regime yields a profound frequency peak at $F$ = 913 T (lower inset of Fig.\,2A).  With magnetic field direction close to the crystal axes, up to four oscillation periods can be seen, as shown in Fig.\,2B.   Also, the overall SdH patterns are almost identical for $\phi$ = 11.3$^\circ$ and $\phi$ = 78.4$^\circ$, and only have a slight valley position shift between $\phi$ = 18.3$^\circ$ and $\phi$ = 70.5$^\circ$, suggesting an axis of symmetry along [101] direction for the SdH oscillations. This is consistent with the cubic structure of YbB$_{12}$ crystal. The fact that the valleys on $d\rho$/$dH$ in Fig.\,2B have approximately constant interval as a function of $1/H$ provides strong evidence of the SdH oscillations with a dominating single frequency.

The observation of the SdH oscillations is reinforced by the $T$-dependence of oscillation amplitudes.  The evolution of the wiggled features in both MR and torque data show typical behaviour of quantum oscillations, with $T$-independent positions of the dominant peaks/valleys within uncertainty and obviously attenuated amplitude from base temperature up to 1.5 K (Fig. S3 in \cite{SM}). The $T$-dependent amplitudes of normalized oscillatory torque (Fig.\,3A) and oscillatory MR (Figs.\,3B and 3C) are fit using the conventional LK formula \cite{shoenberg2009magnetic}. The fittings are reasonably good down to 60\,mK, indicating that the LK theory, established based on the basic Fermi liquid framework, appears to be valid in Kondo insulator YbB$_{12}$.  The agreement with LK description firmly confirms that the features we resolve are quantum oscillation in nature, instead of successive field-induced Lifshitz transitions.

\begin{figure}[t]
\centering
\includegraphics[width=1.0\columnwidth]{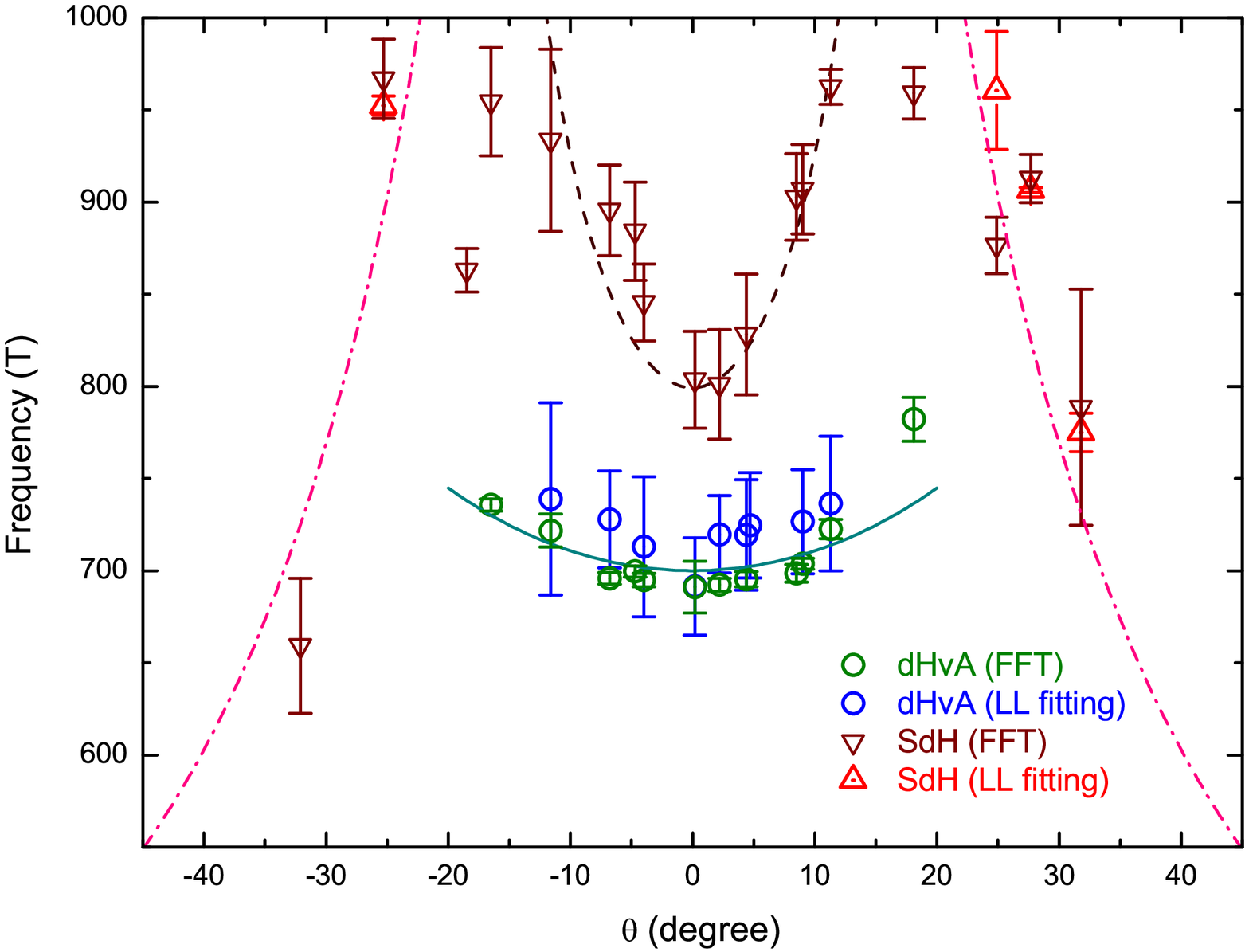}
\caption{{\bf Angular dependence of quantum oscillation frequencies.} The frequencies of the quantum oscillations appearing at high field in YbB$_{12}$. Magnetic field $H$ is rotated in crystalline (010) plane, and the effective tilt angle $\theta$ is defined as the angle between the magnetic field and the equivalent crystal axes [001]/[100] in a cubic structure. Circles are dHvA frequencies obtained from FFT (green) and the slope of linear fitting of LL index versus inverse magnetic field (blue). The dark cyan solid line is a simulation using two-dimensional (2D) Fermi surface model: $F$ = $F_0$/$\cos \theta$ with $F_0$ = 700 T. Up triangles (red) and down triangles (brown) are SdH frequencies acquired by FFT and linear fitting in LL index plot, respectively. The dashed line is fitting by hyperboloid model for a Fermi surface neck region with the principle axis along [001] direction. The error bars come from the difference between different sampling window for FFT results, and the linear fitting error for the LL analysis.
}
\label{FigAngle}
\end{figure}

As shown in Figs.\,3A and 3B, the SdH oscillation is much easier to be suppressed at higher $T$ compared to dHvA oscillations at the same angle, revealing a remarkably heavier effective mass in the electrical transport channel.  The effective masses of the quasiparticles estimated from the dHvA and SdH oscillations at the same tilt angle are approximately 6.6\,$m_e$ and 14.6\,$m_e$, respectively. Therefore it is unlikely that both oscillations originate from the same band.   Indeed the SdH and dHvA frequencies have distinct angle dependence (Fig.\,4).  The dHvA frequencies $F$ can be tracked by a two-dimensional (2D) Fermi surface model ($F \propto \frac{1}{\cos \theta}$) with the in-plane cross section area $A_{[001]}$ = 6.67 nm$^{-2}$ (solid line in Fig.\,4). Given the lack of dHvA oscillations observed above $\simeq$ 20$^\circ$ (Fig.\,S4A in \cite{SM}), this inverse sinusoidal dependence can be explained by either 2D Fermi cylinder, or a heavily elongated 3D Fermi pocket.   On the other hand, the angle dependence of SdH frequencies displays an evident nonmonotonic behaviour: frequency maxima appears at $\theta \sim$ 15-20$^\circ$ from crystal axes, resulting in an ``M"-shape with a local dip at $H \parallel$ [100] and potentially a second minimum at higher tilt angle. The 2D Fermi surface model apparently can not describe this behaviour. Attempt to model the SdH frequencies (dash lines in Fig.\,4, see also Fig.\,S5 in \cite{SM}) points to hyperbolic ``neck" orbits (section S5 in \cite{SM}).


We can rule out a possibility that the bulk SdH oscillations arise from metallic impurity phases. Firstly, as the crystals are grown by floating zone method, additional elements are not included.  The synchrotron X-ray diffraction measurements reveal the absence of metallic Yb-B alloys, including YbB$_2$ and YbB$_4$, within the resolution of 10\,ppm in volume (section S6 and Fig. S6 in \cite{SM}). The only indexed impurity YbB$_6$ shows a much lower dHvA frequency that we observed in YbB$_{12}$ (Fig. S7 in \cite{SM}). Secondly, the angle dependence of both the amplitude and frequency of the observed oscillations coincide with the cubic symmetry of YbB$_{12}$, inconsistent with hexagonal and tetragonal structures of YbB$_2$ and YbB$_4$, respectively  (section S4 and Fig. S5 in \cite{SM}).  Thirdly, according to our simulation assuming the presence of tiny but highly conducting impurities, SdH signal should be detectable at much lower fields (section S7 and Fig. S8 in \cite{SM}), which contradicts with our observations.

The present result is, to our knowledge, the first report of the observation of SdH oscillations in the insulating state.  There are several remarkable features in the SdH oscillations.  The effective masses attained by LK fittings are considerably large, in agreement with the nature of Kondo insulator in which the strong electron correlations make the quasiparticles heavy.  Even in the insulating state, finite specific heat coefficient $\gamma$ is observed in YbB$_{12}$. Interestingly,  $\gamma$ calculated from the SdH oscillations assuming a spherical Fermi surface with $k_F\sim0.17$\,\AA$^{-1}$ ($F = 952$\,T) and effective mass $m^\ast$ = 15 $m_e$, is comparable to the observed value of $\gamma\sim8$\,mJ/mol\,K$^2$ at 40 T \cite{TerashimaThesis}.  The background resistivity $\rho$ still has a magnitude of 10-100 m$\Omega \cdot$cm above 40 T, which is far beyond that of normal metals \cite{Ioffe-Regel}.  Indeed if we estimate the mean free path $\ell$ by considering a spherical Fermi surface with  $m^*$ = 15 $m_e$, then $\rho=0.4\,\Omega\cdot$cm (Fig. 2A) will give a unphysically short mean free path, $\ell \sim 0.01$\,nm, indicating highly unusual situation. A proper theory is yet to be established to describe the quantum oscillations observed in YbB$_{12}$ under high magnetic field (section S8 in \cite{SM}).

The observed unconventional existence of metallic quantum oscillations in insulating ground state reveals that the Kondo insulator is both a charge insulator and a correlated metal, providing a new mysterious and fertile feature of a strongly correlated insulator.\\ \\

\end{document}